\documentclass[lnbip]{svmultln}
\usepackage{makeidx}  
\usepackage[utf8]{inputenc}
\usepackage{hyperref}
\usepackage{algorithm}
\usepackage{algorithmic}
\usepackage{amsmath}
\usepackage{orcidlink}

\usepackage{tabularx}
\usepackage{booktabs}

\usepackage{listings}
\lstdefinestyle{mystyle}{
    language=Java,            
    basicstyle=\small\ttfamily, 
    numbers=left,             
    numberstyle=\tiny,        
    breaklines=true,          
    frame=single,             
		showspaces=false,
		showstringspaces=false, 
    tabsize=4                 
}

\usepackage{pgfplots}
\pgfplotsset{width=7cm,compat=1.15}
\usepgfplotslibrary{statistics}

\usepackage{caption} 
\begin{document}
\mainmatter              
\title{Source Code Clone Detection Using Unsupervised Similarity Measures}

\titlerunning{Unsupervised Source Code Clone Detection}  
%
\author{Jorge Martinez-Gil\orcidlink{0000-0002-5730-7965}}
\authorrunning{Jorge Martinez-Gil}   
%
\tocauthor{Jorge Martinez-Gil}
\institute{Software Competence Center Hagenberg GmbH \\ Softwarepark 32a, 4232 Hagenberg, Austria,\\
\email{jorge.martinez-gil@scch.at}}

\maketitle              

\begin{abstract}        
Assessing similarity in source code has gained significant attention in recent years due to its importance in software engineering tasks such as clone detection and code search and recommendation. This work presents a comparative analysis of unsupervised similarity measures for identifying source code clone detection. The goal is to overview the current state-of-the-art techniques, their strengths, and weaknesses. To do that, we compile the existing unsupervised strategies and evaluate their performance on a benchmark dataset to guide software engineers in selecting appropriate methods for their specific use cases. The source code of this study is available at \url{https://github.com/jorge-martinez-gil/codesim}

\end{abstract}

\section{Introduction}
Source code clone detection holds increasing importance in the current software engineering landscape, and its significance is likely to grow even further \cite{ain2019systematic}. The reason is that this approach is crucial in software development since it can help address various problems during software maintenance \cite{juergens2009code}. Clones are duplicate or similar pieces of code within a software project. Therefore, consider the chaotic situation that would happen if a bug is fixed or a change is made to a piece of code but not to its duplicates. To avoid such situations, developers should have tools to automatically evaluate the likeness between code fragments based on various aspects of their form and functionality \cite{roy2009comparison,roy2007survey}.

In this work, we address this challenge from the point of view of using similarity measures, which are generally used for textual comparisons. When working with general and source code similarity, it is necessary to distinguish between supervised and unsupervised approaches \cite{key-martinez-mlwa}. On the one hand, supervised approaches require a training set of pairs of code fragments labeled as similar or dissimilar, which is often difficult to get, at least in terms of the necessary volume. On the other hand, unsupervised approaches do not require a training set, and they can be used to measure the similarity of any two code fragments with no prior knowledge and a low consumption of computational resources.

This work evaluates at least one representative implementation of unsupervised similarity measures. In this regard, we explore measures ranging from trivial strategies for token comparison to the more advanced comparison of embeddings \cite{key-elmo}. To facilitate a thorough assessment, we use a benchmark dataset comprising diverse code fragments with varying degrees of similarity and check the performance of each similarity measure across the dataset. 

Our analysis focuses on shedding light on practical applicability and efficiency. The rationale behind summarizing the existing body of knowledge and identifying research gaps is to offer a resource for software engineers interested in unsupervised measures for detecting source code clones. Furthermore, in contrast to recent works, which address the challenge from a purely qualitative perspective, our work aims at a quantitative analysis, with an empirical analysis of all the methods considered.

Therefore, this work's primary and overall contribution aims to guide the choice of appropriate unsupervised similarity measures for clone detection. Additionally, it identifies promising directions for future research in source code similarity assessment. The following specific contributions achieve this:

\begin{itemize}
	\item We present the fundamental challenge regarding clone detection and the possibility of building solutions to cope with the absence of labeled data and different coding styles.
	\item We compile an extensive collection of unsupervised semantic similarity measures, being able to compare textual information to elucidate the most promising measures in this context.
	\item We empirically evaluate this collection of unsupervised measures focusing on accuracy, time consumption, practical feasibility, and other metrics such as precision, recall, and f-measures. Our results indicate that several measures could be valid source code clone detection tools.
\end{itemize}

The remainder of this paper is structured as follows: Section 2 introduces the background of this critical challenge of code clone detection. Section 3 technically explains the similarity measures that we are using to face this challenge and shows several examples. Section 4 evaluates all the similarity measures reviewed in the previous version using a complete benchmark dataset. Section 5 discusses the results of our experiments. Finally, the paper concludes with lessons learned and lines of future work.

\section{Background}
This section presents the information necessary to understand the challenge. First, we define code similarity assessment; second, we explain why this challenge is so significant nowadays; and third, we describe the implications and impact of the challenge in academia and industry.

\subsection{Problem definition}
It seems clear that code duplication can lead to inconsistencies, especially if a change is made in one part of the code but not in its clones \cite{novak2019source}. In this context, it is also important to differentiate between code similarity measurement and identification of source code clones. Code similarity measurement is a broad concept, and clone identification is one of its applications. For instance, the most similar instances can be reported as cloned instances just using a threshold value to filter out the results of code similarity measurement \cite{key-baxter}.

Although there is no strict definition for the assessment of code similarity, it is possible to describe the problem formally, such as given a set of code fragments $S = \{C_1, C_2, \ldots, C_n\}$, the goal is to find a function $f: S \times S \rightarrow [0, 1]$ that computes the similarity score between any $C_i$ and $C_j$.

Therefore, $f$ should map a given pair $(C_i, C_j)$ to a value in the continuous interval $[0, 1]$, whereby:

\begin{itemize}
    \item $f(C_i, C_j) = 0$ indicates that $C_i$ and $C_j$ are completely dissimilar
    
    \item $f(C_i, C_j) = 1$ indicates that $C_i$ and $C_j$ are identical
    
    \item $f(C_i, C_j)$ increases as the similarity between $C_i$ and $C_j$ increases and vice versa
\end{itemize}

The function $f$ should compare $C_i$ and $C_j$, considering various characteristics such as variables, constants, function calls, comments, overall logic, or any other code element susceptible to being compared \cite{saini2018code}. Then, clone detection can be implemented to discriminate between instances using, for example, a point value separating clones and non-clones. Furthermore, although it was not considered in the frame of this work, it would be desirable that the results could be accompanied by an explanation \cite{karnalim2021explanation} for facilitating human assessment.

\subsubsection{Similarity categories}
Multiple copies of similar code throughout a software project can make managing the codebase difficult. However, not all the cases are equal. In comparing pieces of code, some recent literature has categorized the code similarities into four categories \cite{aniceto2021source}. These categories help us understand the degree of resemblance between two code fragments so that each category represents a different level of likeness:

\begin{itemize}
	\item Category I: The code fragments are identical, with just minor variations in white spaces and annotations.
	\item Category II: The code fragments have the same structure, but there are differences in the names of the identifiers, data types, spaces, and comments.
	\item Category III: Additionally, parts of the code might be removed or altered, or new parts could be incorporated.
	\item Category IV: The code fragments may appear different but implement analogous functionality.
\end{itemize}

The rationale behind this categorization is to provide insights into code comparison and help software engineers understand the cases they must face to make better-informed decisions. However, more datasets with this categorization are needed, since the existing ones do not usually differentiate.

\subsection{The importance of unsupervised measures}
Detecting code clones is essential for maintaining software quality \cite{higo2002software}. Unsupervised code similarity assessment can help address this challenge since several practical aspects are common to many software development projects:

\begin{itemize}
	\item Unsupervised measures do not rely on labeled training data, making use of a ground truth unnecessary. Labeled examples are only needed to validate the performance of unsupervised approaches.

	\item Code can be written in various programming languages, using different coding styles, etc. Some unsupervised measures can accommodate this variety without a complex universal similarity metric.

	\item Understanding the meaning of code is complicated because code fragments may be functionally equivalent even if they look dissimilar, and vice versa. Some unsupervised measures can face that challenge.

	\item Codebases often contain comments, noise, etc. Some unsupervised measures can differentiate between meaningful code patterns and unrelated elements.
\end{itemize}

\subsection{Future perspectives}
Duplicate code increases the maintenance burden because changes must be replicated across all clones, which is time-consuming and error-prone \cite{bellon2007comparison}. Therefore, identifying and refactoring these clones can reduce the maintenance effort \cite{dou2023towards}. 

Nowadays, where many open-source libraries and code repositories exist, unsupervised source code similarity measurement can be helpful; it enables developers to navigate this diverse ecosystem and search for relevant code efficiently with low consumption of computational resources \cite{key-martinez-kbs}. This importance extends to facilitating code reuse, which is crucial for reducing development time in the face of growing software complexity \cite{dang2011code}. 

Furthermore, detecting code similarities can improve security by identifying vulnerabilities with known code patterns in the context of growing security troubles. It can also contribute to code maintainability and refactoring efforts, allowing developers to ensure software projects' long-term sustainability.

We can also think of applications within various industries that benefit from increased compliance and reliability in critical systems. Furthermore, collaboration tools facilitate cooperation by connecting developers with similar code, and quality assurance strategies could benefit from unsupervised code similarity measurement by identifying similar cases for complete test coverage.

\section{Methods}
Early approaches for assessing the similarity relied on just textual analysis \cite{key-martinez-jiis}. These techniques, while efficient, often struggle to capture the structural aspects of code, resulting in limited accuracy \cite{ferrante1987program}. However, the field has evolved a lot in recent years. More sophisticated similarity measures assumed to perform better have been proposed \cite{key-Bert}.

\subsection{Unsupervised methods}
There are many methods (a.k.a. semantic similarity measures) to determine the similarity between textual entities. Each measure offers a unique approach based on specific characteristics or representations of the compared entities. From the literature, we have identified about 21 families that could be applied here, briefly explained below in alphabetical order.

\begin{itemize}
    \item \textbf{Abstract Syntax Trees (ASTs) Similarity}: ASTs are hierarchical representations of the structure of code. AST similarity measures compare the structural similarity between different AST representing code \cite{karnalim2020syntax}.
    \item \textbf{Bag-of-Words Similarity}: This similarity measure calculates the resemblance between texts by considering the frequency of individual words in each text without considering word order or structure \cite{corley2005measuring}.
    \item \textbf{Code Embeddings Similarity (CodeBERT)}: Code embeddings are vector representations of source code. This method measures the similarity of code based on these embeddings \cite{alon2019code2vec}. Please note that we use them here without recalibration.
    \item \textbf{Comments Similarity}: It measures the similarity between code comments, which can be helpful for code documentation and understanding. In principle, many traditional text similarity measures can be used \cite{key-martinez-eswa2}.
    \item \textbf{Fuzzy Matching Similarity}: Fuzzy matching compares strings for minor syntactical variations. It is often used in data matching and search applications \cite{singla2012string}, but we apply it here to measure code similarity.
    \item \textbf{Function Calls Similarity}: This family measures the similarity between different code fragments based on the functions and procedures in the code fragments \cite{xu2013similarity}.
    \item \textbf{Graph-based Similarity}: It calculates similarity based on a graph's relations, which could represent various data structures and dependencies \cite{zager2008graph}.
    \item \textbf{Jaccard Similarity}: Jaccard similarity measures the similarity between sets of tokens by comparing their intersection and union. It is commonly used in text analysis, recommendation systems, and information retrieval \cite{haque2022semantic}.
    \item \textbf{Levenshtein Similarity}: This measure, also known as edit distance, calculates the similarity between two strings by measuring the number of edits needed to transform one into the other \cite{key-levenshtein}.
    \item \textbf{Longest Common Subsequence (LCS) Similarity}: LCS similarity calculates the similarity between two sequences by finding the longest common subsequence between them \cite{bergroth2000survey}.
    \item \textbf{Metrics Similarity}: The idea is first to compute various metrics related to the source code and then estimate the similarity between the values obtained \cite{nunez2017source}. We are using here: code length, cyclomatic complexity, number of variables, etc.
    \item \textbf{N-grams Similarity}: N-grams are contiguous sequences of 'n' items (e.g., words or characters). N-gram similarity measures the similarity between texts based on shared n-grams \cite{damashek1995gauging}.
    \item \textbf{Output Analysis Similarity}: This method measures the similarity of program outputs, which can be helpful for testing and debugging. In principle, and if we assume the outputs as text, a wide range of traditional text similarity measures can be used \cite{key-martinez-eswa2}.
    \item \textbf{Perceptual Hashing Similarity}: Perceptual hashing, often used in image similarity, aims to generate a fixed-length hash code from images. In our context, this method measures similarity based on hashes from visual representation of the code \cite{key-Ragkhitwetsagul}.
    \item \textbf{Program Dependence Graph Similarity}: This measure assesses the similarity between code by analyzing the program dependence graph, which represents the dependencies between program elements \cite{krinke2001identifying}. It is different from the Graph-based method since focuses on control dependencies.
    \item \textbf{Rolling Hash Similarity}: A rolling hash is a hash function that can be updated efficiently as new data is processed. Rolling hash similarity can compare substrings (hashes) in large texts \cite{hartanto2019best}. We use here for comparing code.
	\item \textbf{Running-Karp-Rabin Greedy-String-Tiling (RKR-GST) Similarity}: It is often used in the context of detecting plagiarism by identifying maximal sequences of contiguous matching tokens (tiles) \cite{wise1993string}.
    \item \textbf{Semdiff Similarity}: Semdiff is a method for detecting semantic differences between program versions. Semdiff similarity measures how code changes affect the program's semantics \cite{horwitz1990identifying}.
    \item \textbf{Semantic Clone Similarity}: This method family tries to measure the similarity of code fragments based on the semantic meaning of the names of the program elements (variables, methods, etc.) \cite{gabel2008scalable}.
    \item \textbf{TF-IDF Similarity}: Term Frequency-Inverse Document Frequency (TF-IDF) is used in text analysis to measure the importance of words in a text compared to a larger corpus. TF-IDF similarity compares texts based on these weighted terms \cite{karnalim2020tf}.
    \item \textbf{Winnow Similarity}: It is a text comparison algorithm that identifies similar texts by hashing them and comparing their fingerprints \cite{schleimer2003winnowing}.
\end{itemize}

Next, we will look at some Java code examples, representing some interesting cases of source code cloning, illustrating how all these similarity measures quantify code similarity in practice.

\subsection{Examples}
In the examples below, \textit{T1} and \textit{T01} are two Java classes that produce the same output but with different approaches: \textit{T1} prints the statement \textit{Welcome to Java} five times using five separate print statements. \textit{T01} achieves the same output using a for loop that iterates five times, printing the statement on each iteration. From the perspective of code clone detection, these two classes are Category IV clones. The reason is that both are pieces of code that perform the same operations but are implemented through different syntactic variations. 

Even though the actual text of the code differs, the for loop versus repeated print statements, the meaning, and the output are the same. However, detecting such code clones can be challenging because it is not just a matter of matching text strings but requires a deep understanding of the code's logic. However, it is common to find similar cases in real settings.

\begin{lstlisting}[style=mystyle]
public class T1 {
    public static void main(String[] args) {
        System.out.println("Welcome to Java");
        System.out.println("Welcome to Java");
        System.out.println("Welcome to Java");
        System.out.println("Welcome to Java");
        System.out.println("Welcome to Java");
    }
}
\end{lstlisting}

\begin{lstlisting}[style=mystyle]
public class T01 {
    public static void main(String[] args){
        
        for(int i = 0; i < 5; i++){
            System.out.println("Welcome To Java");
        }
        
    }
}
\end{lstlisting}

On the contrary, the classes \textit{TemperatureConverter} and \textit{CurrencyConverter} are similar in form. However, an experienced developer would quickly realize that they calculate different things (temperature vs currencies), so they should not be considered clones. However, their high similarity in form might make many unsupervised measures consider them Category II clones.

\begin{lstlisting}[style=mystyle]
public class TemperatureConverter {
    public static double celsiusToFahrenheit(double cels) {
        return cels * 9 / 5 + 32;
    }
}
\end{lstlisting}

\begin{lstlisting}[style=mystyle]
public class CurrencyConverter {
    public static double usdToEur(double usd) {
        return usd * 85 / 100;
    }
}
\end{lstlisting}

Table \ref{tab:ac} compares various unsupervised similarity measures for code analysis. Some of these measures are based on textual similarity, while others are based on the structure of the code. Other measures might analyze the code's functionality beyond just the text or structure. In principle, there is no accurate or inaccurate result in this context. However, intuition tells us that some measures may better serve our purposes. The ideal result would be 1.00 in the first column and 0.00 in the second. Nevertheless, any result that can discern clones (giving them a high similarity value) from non-clones (giving them a low similarity value) would be good.

\begin{table}[h]
\centering
\begin{tabularx}{\textwidth}{lcc}
\toprule
\textbf{Measure} & \textbf{Score-Ex1.} & \textbf{Score-Ex2.} \\
\midrule
Abstract Syntax Trees (ASTs) Similarity & 0.50 & 0.81 \\
Bag-of-Words Similarity & 0.72 & 0.65 \\
Code Embeddings Similarity & 0.99 & 1.00 \\
Comments Similarity & 1.00 & 1.00 \\
Fuzzy Matching Similarity & 0.54 & 0.64 \\
Function Calls similarity & 1.00 & 0.00 \\
Graph-based Similarity & 0.38 & 0.34 \\
Jaccard Similarity & 0.27 & 0.35 \\
Levenshtein Similarity & 0.51 & 0.69 \\
Longest Common Subsequence (LCS) Similarity & 0.19 & 0.29 \\
Metrics Similarity & 0.98 & 1.00 \\
N-grams Similarity & 0.26 & 0.14 \\
Output Analysis Similarity & 1.00 & 0.00 \\
Perceptual Hashing Similarity & 0.69 & 0.88 \\
Program Dependence Graph Similarity & 1.00 & 1.00 \\
R.-Karp-Rabin G.-Str.-Til. (RKR-GST) Similarity & 0.96 & 0.83 \\
Rolling Hash Similarity & 1.00 & 0.55 \\
Semdiff Similarity & 0.22 & 0.40 \\
Semantic Clone Similarity & 0.54 & 0.79 \\
TF-IDF Similarity & 0.67 & 0.48 \\
Winnow Similarity & 1.00 & 0.60 \\
\bottomrule
\end{tabularx}
\caption{Comparison of various unsupervised similarity measures for code similarity measurement}
\label{tab:ac}
\end{table}

Please note that something special happens with the \textit{Comments Similarity} result. Since none of the displayed code fragments have comments, the measure thinks they are similar. This is just an example of why caution is necessary when considering the results.

\section{Evaluation}
Several aspects come into play when evaluating and comparing unsupervised similarity measures for clone detection. To effectively evaluate these techniques, it is essential to consider the dataset's nature, the clone categories to face, and the task's requirements. 

In this way, some measures excel in comparing textual content, making them suitable for detecting cloned text. Other techniques are more apt for identifying similar functionality. In contrast, other measures can assist in uncovering structural similarities between code and text. The choice depends on the nature of the data in the benchmark dataset.

\subsection{Dataset}
We are using here the IR-Plag dataset\footnote{\url{https://github.com/oscarkarnalim/sourcecodeplagiarismdataset}} which is designed to serve as a benchmark for evaluating and comparing the performance of different strategies \cite{key-karnalim}. This dataset includes plagiarized code files deliberately crafted to mimic academic plagiarism behaviors. Although the dataset is compiled to detect plagiarism, it is valid for our purposes since the practical result of plagiarism and cloning is the same in practice, even if their original intentionality might differ (intention to deceive in the first case, no intentionality in the second). Moreover, this dataset does not merely focus on simplistic plagiarism attacks but encompasses a complete range of complexities. Although this dataset does not classify clones, it can be useful in detecting suitable semantic similarity measures for mitigating code redundancy and duplication within complex software projects. 

In analyzing a dataset of code files, we observe the following metrics: The dataset contains seven original code files (original programming assignments). A high number of files, 355 (77\%), are identified as plagiarized, suggesting a considerable prevalence of duplication. There are 105 non-plagiarized files, which might represent modified or derivative works. The total count of code files in the dataset is 467. Within these files are 59,201 tokens, with 540 distinct tokens, indicating the variety of programming language elements used. The size of the files varies significantly, with the largest file containing 286 tokens and the smallest comprising 40 tokens. On average, a code file in this dataset includes around 126 tokens. These insights show the dataset's composition, reflecting a great diversity in programming syntax.

\subsection{Results}
In the following, we show the results obtained from the experiments on the IR-Plag dataset. We look primarily at the accuracy (hit percentage) and the execution time required as we believe these are two of the most important aspects to consider when considering putting a measure into operation. These results can be reproduced with the provided source code\footnote{\url{https://github.com/jorge-martinez-gil/codesim}}.

\begin{figure}[h]
\centering
\includegraphics[width=\textwidth]{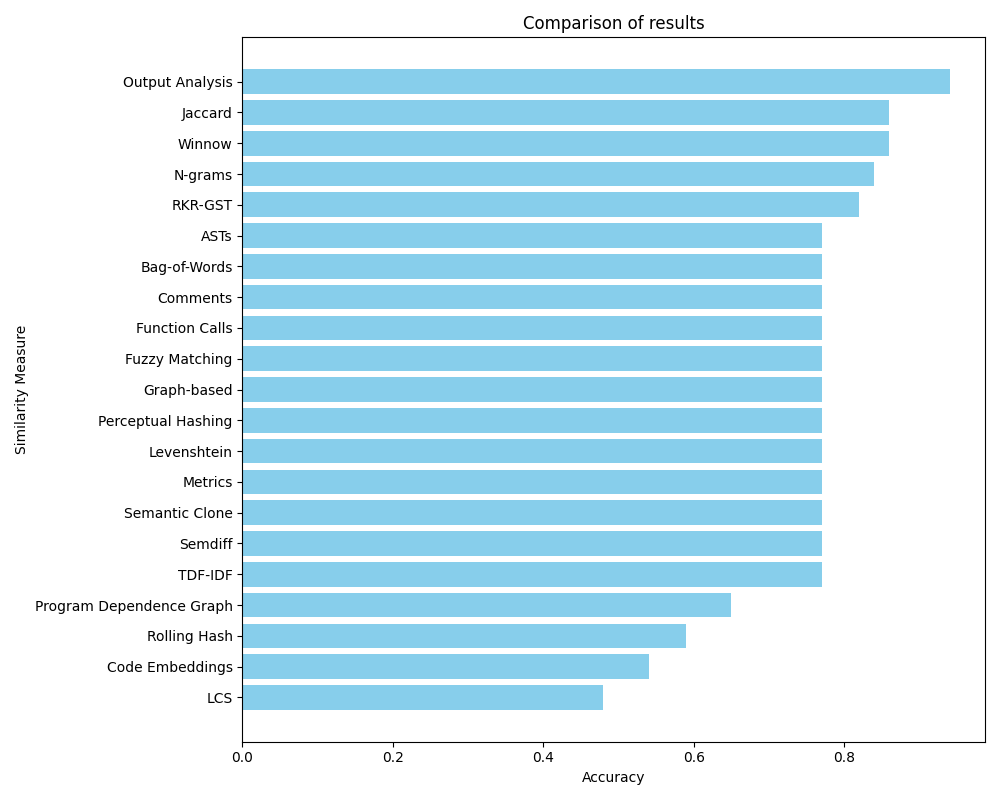}
\caption{Accuracy of the unsupervised semantic similarity measures when performing clone detection}
\label{fig:acc}
\end{figure}

On the one hand, Figure \ref{fig:acc} compares the different measures. The horizontal axis quantifies the accuracy of each measure, while the vertical axis lists the unsupervised measures. \textit{Output Analysis} has the highest accuracy score, which could imply that it is most effective at detecting code that performs the same function despite differences in implementation. Contrariwise, \textit{LCS} has the lowest accuracy score, indicating that it might not be as effective in this comparison.

It is essential to note that the dataset contains 77\% clones. Therefore, a simplistic approach could be to classify all comparisons as clones, which would result in achieving an accuracy of 0.77 by default. This would not be a good result. Figure \ref{fig:acc} shows that only using 5 measures produces a real gain over that base result.

On the other hand, Figure \ref{fig:vel} presents a comparative analysis of various measures used to execute code, measured by their execution time. The horizontal axis quantifies the execution time, while the vertical axis lists the unsupervised measures. The \textit{Output Analysis} shows the longest execution time, significantly outpacing other methods such as \textit{Comments} and \textit{Code Embeddings}. The remaining measures show lower execution times, suggesting a more efficient performance. Two facts can be immediately deduced from these experiments: 

\begin{figure}[h]
\centering
\includegraphics[width=\textwidth]{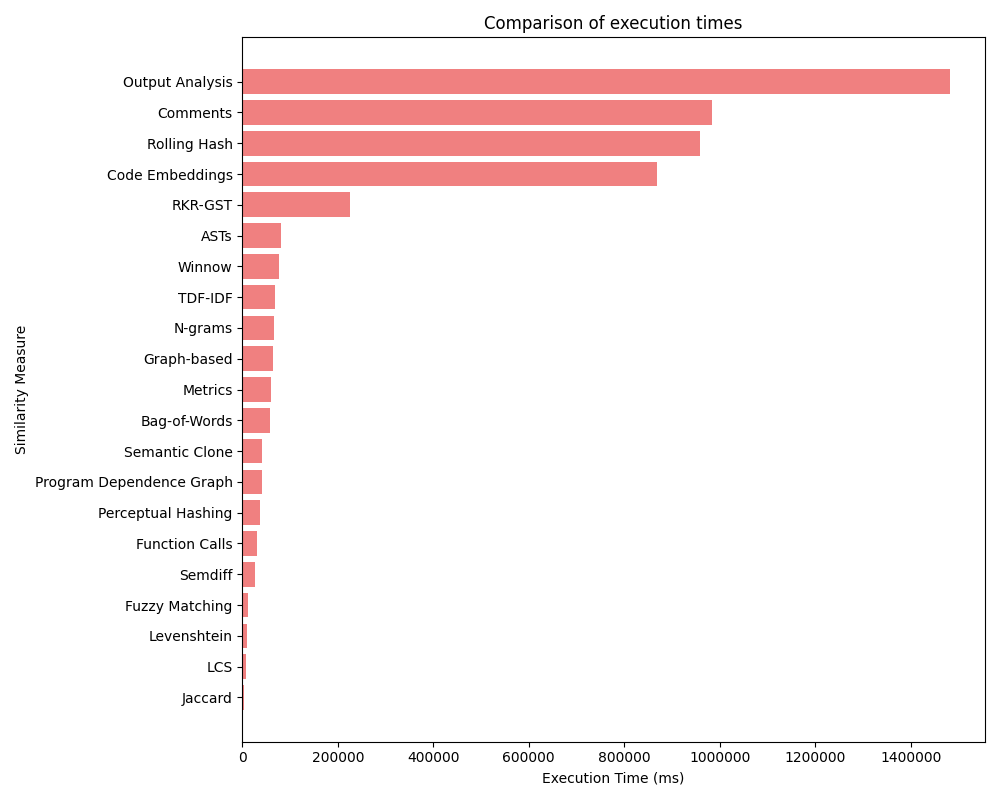}
\caption{Execution time of the unsupervised semantic similarity measures when performing clone detection}
\label{fig:vel}
\end{figure}

\begin{enumerate}
	\item First, only five of the measures studied (i.e., \textit{Output Analysis}, \textit{Winnow}, \textit{N-grams}, \textit{RKR-GST}, and \textit{Jaccard}) help identify clones effectively. This suggests that most unsupervised semantic similarity measures are not helpful in the current form. Therefore, more research on innovative approaches to clone detection is needed. 
	\item  Despite being excellent in accuracy (e.g., \textit{Output Analysis}), some techniques incur such a high computational cost that incorporating them into a practical, real-world tool for programmers becomes unrealistic. The reason \textit{Output Analysis} takes so much execution time is that it must take the two pieces of code, encapsulate them for compilation, pass some random parameters to them (if necessary), and compare the outputs produced. This entire process is very computationally expensive.
\end{enumerate}
 
Other time-consuming similarity measures are \textit{Rolling Hash} (very intensive in the use of mathematical operations), \textit{Comments Similarity} (identifying comments involves the use of regular expressions, which is computationally expensive), and \textit{Code Embeddings} (which needs to search and identify embeddings as well as perform operations on them). Therefore, it would be possible to define a feasibility index that calculates a combination of accuracy and execution time to elucidate which measures could work well in real environments. This could be done by weighing the importance of accuracy about time and dividing the result by the total execution time.

\begin{figure}[h]
\centering
\includegraphics[width=\textwidth]{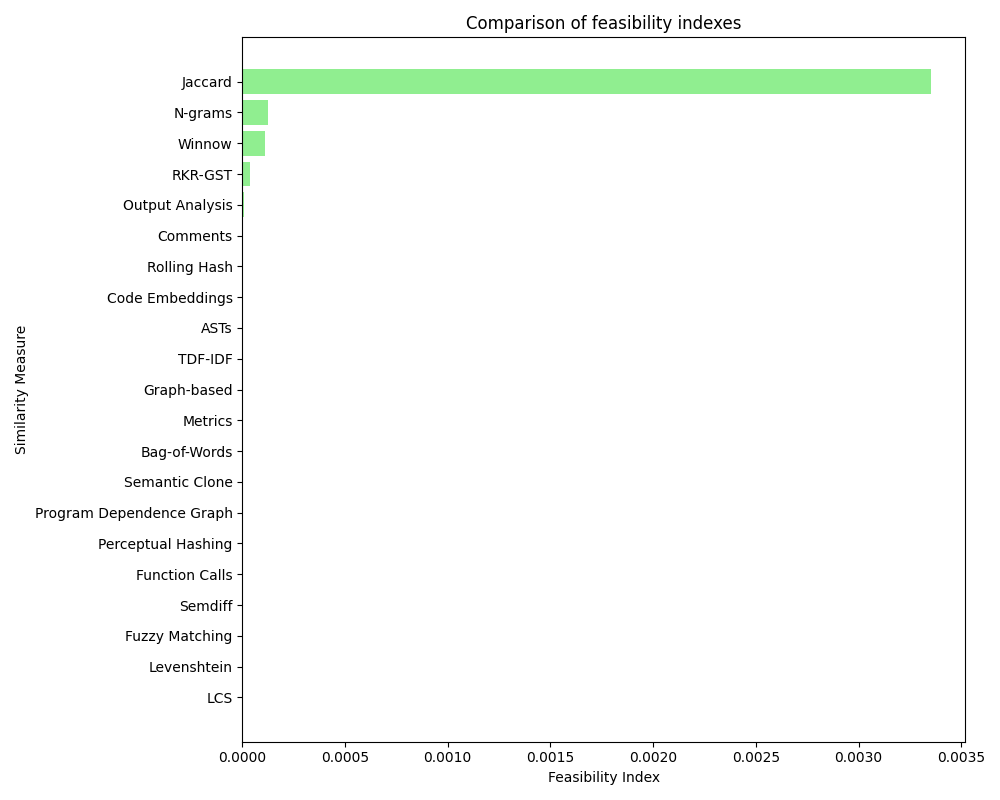}
\caption{Comparison of the feasibility index of the unsupervised methods}
\label{fig:fea}
\end{figure}

Figure \ref{fig:fea} shows us the calculation of the feasibility index. We consider the accuracy importance over the execution time as 10:1. Therefore, just \textit{Jaccard}, \textit{N-grams}, \textit{Winnow}, and \textit{RKR-GST}(in that order) would be good candidates for use in real environments due to a reasonable combination of accuracy and execution time. However, these measures should be used just for an automatic recommendation since the gain in accuracy over the base result only allows us to operate them with supervision.

\subsection{Other metrics}
Apart from accuracy, there are other metrics from the information retrieval field to assess how well a clone detection system performs in terms of accuracy and completeness. 

\begin{itemize}
	\item Precision: The approach's accuracy is evaluated by measuring the proportion of correctly identified true clones among all identified code fragments. Higher precision means greater reliability.
	\item Recall: Recall, also known as sensitivity, assesses the approach's completeness. It quantifies the proportion of true clones in the dataset that were successfully identified. A higher recall indicates fewer missed clones.
	\item F-measure: The F-measure evaluates overall performance by balancing precision and recall. It ensures a well-rounded assessment of both precision and recall.
\end{itemize}

This way of evaluating is also popular since it gives more weight to the positive classes by considering false positives (precision) and false negatives (recall) separately. It penalizes the model for failing to detect positive cases and making false positive predictions.

Table \ref{tab:clonemetrics} shows us the results that can be obtained for these metrics from the information retrieval field with the unsupervised similarity measures that we have been using throughout this work.

\begin{table}[h]
\centering
\begin{tabular}{lccc}
\toprule
Measure & Precision & Recall & F-Measure \\
\midrule
Abstract Syntax Trees (ASTs) Similarity & 0.77 & 0.78 & 0.78 \\
Bag-of-Words Similarity & 0.79 & 0.66 & 0.72 \\
Code Embeddings Similarity & 0.75 & 0.34 & 0.47 \\
Comments Similarity & 0.77 & 1.00 & 0.87 \\
Function Calls similarity & 0.78 & 0.91 & 0.84 \\
Fuzzy Matching Similarity & 0.77 & 1.00 & 0.87 \\
Graph-based Similarity & 0.80 & 0.52 & 0.63 \\
Jaccard Similarity & 0.81 & 0.94 & 0.87 \\
Levenshtein Similarity & 0.80 & 0.66 & 0.72 \\
Longest Common Subsequence (LCS) Similarity & 0.74 & 0.06 & 0.11 \\
Metrics Similarity & 0.77 & 1.00 & 0.87 \\
N-grams Similarity & 0.84 & 0.29 & 0.43 \\
Output Analysis Similarity & 0.85 & 0.97 & 0.90 \\
Perceptual Hashing Similarity & 0.77 & 0.85 & 0.81 \\
Program Dependence Graph Similarity & 0.85 & 0.39 & 0.53 \\
R.-Karp-Rabin G.-Str.-Til. (RKR-GST) Similarity & 0.79 & 0.99 & 0.88 \\
Rolling Hash Similarity & 0.93 & 0.18 & 0.30 \\
Semdiff Similarity & 0.79 & 0.38 & 0.51 \\
Semantic Clone Similarity & 0.79 & 0.68 & 0.73 \\
TF-IDF Similarity & 0.77 & 0.99 & 0.87 \\
Winnow Similarity & 0.81 & 0.98 & 0.88 \\
\bottomrule
\hline
\end{tabular}
\caption{Comparison of the unsupervised semantic similarity measures using other popular metrics}
\label{tab:clonemetrics}
\end{table}

As can be seen, the \textit{Code Embeddings Similarity} approach stands out with the highest precision, indicating its remarkable accuracy in identifying code clones. For comprehensive clone detection, \textit{Output Analysis Similarity} and \textit{Program Dependence Graph Similarity} excel with good recall values, implying their ability to capture a significant portion of true clones. We exclude \textit{Comments Similarity} for reasons already commented about the dataset's low importance of comments. 

If we look for a balanced performance that combines precision and recall, \textit{Output Analysis Similarity} offers an attractive option, boasting the highest F-Measure at 0.90. Exactly as was the case with accuracy; however, its high execution times would still give it no option in exploitation environments, so \textit{Jaccard}, \textit{RKR-GST}, and \textit{Winnow} would be, again, more suitable. In this occasion, \textit{N-grams} should not be considered due to its low recall.

\section{Discussion}
Our experiments show that a reduced group of unsupervised source code similarity measurements could be used to detect source code clones. These methods could improve various aspects of software engineering. For example, they could suggest the existence of clones and, therefore, present an opportunity to refactor the code into reusable parts. This might facilitate code reuse, which is vital in software engineering. 

It is also necessary to remark that noisy and unstructured code environments characterize real-world computing environments. We have identified several unsupervised similarity measures that have shown promise in managing this noise and variability, making them valuable when labeled data is limited or impractical. However, the majority of similarity measures studied would not be suitable for this purpose.

Despite some progress, our research still needs to solve several challenges. These include achieving cross-language similarity measurement and ensuring scalability for large codebases. These challenges present compelling opportunities for future research. This means that our research results, although slightly applicable in their current form, need further research to be useful as software development advances into a more automated future.

\section{Conclusion}
The challenge of source code clone detection represents a very important aspect of software engineering that impacts many diverse applications. In this work, we have evaluated the existing unsupervised similarity measures to address the challenges of the absence of labeled data and diverse coding styles. Our research illustrates how unsupervised source code similarity measurement can facilitate clone detection. 

As codebases grow, developing accurate and efficient unsupervised similarity measures remains an essential area of exploration for the community. Furthermore, the need for effective unsupervised techniques will likely expand as the software industry evolves. Although studying supervised techniques may promise good results, unsupervised techniques will always be an option due to their more realistic requirements, adaptability, interpretability, and efficiency.

Therefore, the future of source code clone detection using unsupervised measures holds notable promise. Future efforts could focus on hybrid approaches that integrate the strengths of different methods (a.k.a. ensembles), leading to more robust and accurate similarity assessments. Exploring transfer learning techniques could also improve performance. The goal should be to enhance strategies for code analysis with good accuracy but minimal human intervention.

\section*{Acknowledgments}
The author thanks all the anonymous reviewers for their help in improving the manuscript. The research reported in this paper has been funded by the Federal Ministry for Climate Action, Environment, Energy, Mobility, Innovation and Technology (BMK), the Federal Ministry for Labour and Economy (BMAW), and the State of Upper Austria in the frame of the SCCH competence center INTEGRATE [(FFG grant no. 892418)] in the COMET - Competence Centers for Excellent Technologies Programme managed by Austrian Research Promotion Agency FFG.

\bibliographystyle{plain}
\bibliography{mybib}
\end{document}